\documentstyle[]{mn}

\input epsf
\input psfig.sty

\renewcommand{\d}{{\rm d}}

\newcommand{\beq}{\begin{equation}}
\newcommand{\eeq}{\end{equation}}
\newcommand{\beqa}{\begin{eqnarray}}
\newcommand{\eeqa}{\end{eqnarray}}
\newcommand{\bea}{\begin{array}}
\newcommand{\ea}{\end{array}}
\newcommand{\cG}{{\cal G}}
\newcommand{\rhob}{\overline{\rho}}
\newcommand{\lag}{\langle}
\newcommand{\rag}{\rangle}
\newcommand{\Om}{\Omega_{\rm m}}
\newcommand{\Ol}{\Omega_{\Lambda}}
\newcommand{\Ob}{\Omega_{\rm b}}
\newcommand{\xib}{\overline{\xi}}
\newcommand{\tcool}{t_{\rm cool}}

\newcommand{\dR}{\delta_R}

\newcommand{\LX}{L_{\rm X}}
\newcommand{\cP}{{\cal P}}
\newcommand{\nb}{n_{\rm b}}
\newcommand{\Sw}{S_{\rm w}}
\newcommand{\Tg}{T_{\rm g}}
\newcommand{\Ts}{T_{\rm s}}
\newcommand{\Tad}{T_{\rm ad}}
\newcommand{\Tvir}{T_{\rm vir}}

\newcommand{\rhog}{\rho_{\rm g}}
\newcommand{\Rcore}{R_{\rm core}}
\newcommand{\Deltavir}{\Delta_{\rm vir}}
\renewcommand{\ng}{n_{\rm g}}
\newcommand{\Svir}{S_{\rm vir}}
\newcommand{\Ssub}{S_{\rm sub}}
\newcommand{\Sex}{S_{\rm ext}}

\newcommand{\Fvir}{F_{\rm vir}}

\title[Are clusters born warm?]{The phase-diagram of the IGM and the entropy floor of groups and clusters: are clusters born warm?}   
\author[P. Valageas, R. Schaeffer and J. Silk]{Patrick Valageas $^1$, Richard Schaeffer $^1$ and Joseph Silk $^2$\\
$^1$Service de Physique Th\'eorique, CEN Saclay, 91191 Gif-sur-Yvette, France\\
$^2$Astrophysics, Department of Physics, Keble Road, Oxford OX1 3RH, UK} 

\begin{document}

\maketitle

\begin{abstract}
We point out that two problems of observational cosmology, the facts i) that $\ga 60 \%$ of the baryonic content of the universe is not observed at $z\sim 0$ and ii) that the properties of small clusters do not agree with simple expectations, could be closely related. As shown by recent studies, the shock-heating associated with the formation of large-scale structures heats the intergalactic medium (IGM) and leads to a ``warm IGM'' component for the gas. In the same spirit, we suggest the intracluster medium (ICM) to be a mixture of galaxy-recycled, metal enriched gas and intergalactic gas, shock-heated by the collapsing much larger scales. This could be obtained through two processes: 1) the late infalling gas from the external warm IGM is efficiently mixed within the halo and brings some additional entropy, or 2) the shocks generated by larger non-linear scales are also present within clusters and can heat the ICM. We show that if assumption (1) holds, the entropy brought by the warm IGM is sufficient to explain the observed properties of clusters, in particular the entropy floor and the $\LX-T$ relation. On the other hand, we briefly note that the scenario (2) would require a stronger shock-heating because of the larger density of the ICM as compared with filaments. Although the efficiency of these two processes remains to be checked on a quantitative level, they present the advantage to dispense with the need to invoke any strong preheating from supernovae or quasars (which has otherwise been introduced for the sole purpose of reproducing the behaviour of clusters). Matter ejection by galaxies is included in the present calculations, and consistently with the metal-enrichment requirements, is indeed shown to yield only a quite moderate entropy increase. Our scenario of clusters being "born warm" can be checked through the predicted redshift evolution of the entropy floor.
\end{abstract}


\section{Introduction}

A key test of cosmological scenarios is to check whether they can reproduce the large-scale structure of the present universe. Many studies have already shown that the usual hierarchical scenarios (such as  the standard CDM model) agree reasonably well with observations of galaxies (Valageas \& Schaeffer 1999, Kauffmann et al. 1993, Cole et al. 1994), Lyman-$\alpha$ clouds (Valageas et al. 1999, Petitjean et al. 1992, Miralda-Escude et al. 1996), and quasars, as well as with  reionization constraints (Valageas \& Silk 1999a, Gnedin \& Ostriker 1997, Haiman \& Loeb 1998). 

However some puzzles still remain unresolved. For instance, the mass of baryons observed at $z=0$ (stars, intracluster hot gas) yields $\Ob \sim 0.01$ (Fukugita et al. 1998), while the Lyman alpha forest accounts for about as much (Penton et al. 2000). This falls short of the CMB determination by Archeops (Benoit et al. 2003), $\Ob \simeq 0.052 \pm 0.007$, or the prediction of standard nucleosynthesis $\Ob \simeq 0.045 \pm 0.006$ (Tytler et al. 2000) \footnote{Our choice of cosmological parameters is given in Section \ref{Phase-diagram of cosmological baryons}.}. As argued in Cen \& Ostriker (1999) and Dave et al. (2001), up to $50\%$ of the baryons could be in the form of a ``warm" phase of the intergalactic medium (IGM), with temperatures in the range $10^5 < T <10^7$ K, and provide the explanation for the ``missing" baryonic matter.

A second problem is the temperature - X-ray luminosity relation of clusters. Indeed, the simplest models yield $\LX \propto T^2$ (Kaiser 1986) which disagrees with observations (Ponman et al. 1996). It has been suggested (e.g., Evrard \& Henry 1991, Cavaliere et al. 1997, Ponman et al. 1999) that this discrepancy could be due to a ``preheating'' of the IGM which would raise its entropy before clusters form. Such an entropy ``floor'' breaks the standard scaling laws and  yields a steeper  $T - \LX$ relation, and may possibly  play a  role in the solution of the ``overcooling'' problem (Blanchard et al. 1992). Two sources of energy which could provide this preheating appear naturally in the usual cosmological scenarios: supernovae/stellar winds and quasars. However, realistic models (e.g., Valageas \& Silk 1999b, Wu et al. 2000, Bower et al. 2001) show that the energy released by supernovae is unlikely to be sufficient (unless the energy transfer is highly efficient and targeted to the gas which will build clusters and groups). On the other hand, the energy provided by quasars is largely  sufficient (Valageas \& Silk 1999b), but it is not easy to estimate with a good accuracy the efficiency of the energy transfer. Although it is commonly agreed that an increase in entropy is required, its actual origin remains an open issue.

Some numerical simulations (Muanwong et al. 2001), on the other hand,  show that inhomogeneous cooling of the baryonic gas, which removes the cooler gas as it condenses into galaxies while leaving only the hotter component available as intracluster gas, can play the same role as an increase in entropy. A related proposal (Voit \& Bryan 2001) is that the entropy along the edge of the cooling curve matches, at a given virial temperature, the observed entropy of the intracluster gas. However, to be efficient this may imply an excessive amount of cooling (Muanwong et al. 2001).

In this work, we investigate another source of entropy: the shock-heating associated with the gravitational collapse of very large scales which are just turning non-linear. This is partly motivated by the recent finding in numerical simulations of a warm IGM phase generated by such a {\it transfer of energy from collapsing very large scales to much lower scales}, see Dave et al. (2001). This mechanism, driven by the non-linear length scale ($\sim 9$ Mpc at $z=0$), heats the IGM which is clustered within much smaller filaments (i.e. with a similar length but a smaller thickness of order of a few hundred kpc at $z=0$) and occupies a small fraction of the volume of the universe (e.g., Cen \& Ostriker 1999, Valageas et al. 2002). Since clusters are located at the crossing of such filaments, their properties could also be influenced by this shock-heating. Therefore, in this study we investigate whether the entropy floor observed within clusters could be generated by this shock-heating from scales much above the cluster scale. This could occur through two mechanisms. First, the warm IGM embedded within neighbouring filaments, where it underwent this shock-heating, could be efficiently mixed within the cluster as the accretion which builds the halo proceeds. Then, the additional entropy brought with this flow of matter could increase the entropy of the cluster. In order to be relevant, this process requires a sufficiently efficient mixing of the gas so that at least the small clusters with a temperature below $\sim 1$ keV which display the anomaly are affected in their bulk gas distribution. We shall evaluate the level of mixing required by such a scenario. A second mechanism would be for the gas to be shock-heated {\it in situ}, that is the shocks generated by the large scales which are just turning non-linear may run through the cluster and heat the ICM itself. We shall not study this alternative scenario in detail here, but we note that it requires stronger shocks in order to significantly increase the entropy of this higher-density gas.

\section{Phase-diagram of cosmological baryons}
\label{Phase-diagram of cosmological baryons}

For consistency with our earlier work, we use a Hubble constant $H_0 = 65$ km s$^{-1}$, and consider a standard $\Lambda$CDM universe with $\Om=0.4$ and $\Ol=0.6$, with a baryonic density parameter of $\Ob=0.047$. The $\Lambda$CDM power-spectrum of the linear density fluctuations is normalized by $\sigma_8=0.8$.

\begin{figure}
\psfig{figure=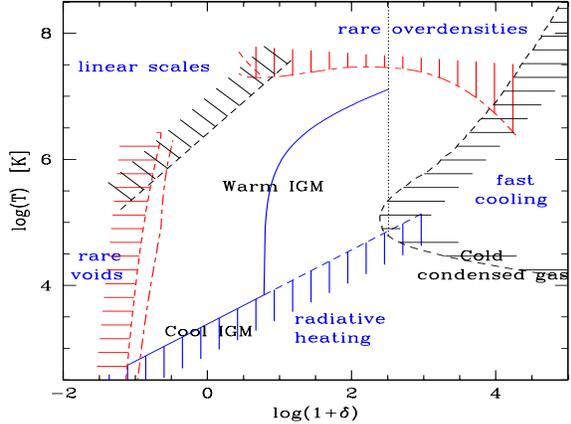,width=8cm,height=6cm}
\caption{The phase-diagram of cosmological baryons at $z=0$ in a $\Lambda$CDM universe. We can distinguish two IGM phases: the ``cool'' IGM (Lyman-$\alpha$ forest), which follows a well-defined equation of state, and the ``warm'' IGM, which is constrained to lie within the central domain delimited by the exclusion regions represented by the hatching. The vertical dotted line at $\delta \sim 200$ shows the density contrast of just-virialized halos.}
\label{diagz0} 
\end{figure}

First, we review in Fig.\ref{diagz0} the properties of the IGM at redshift $z=0$, as obtained in Valageas et al. (2002). These can be described by the distribution of the baryons within the $(\rho,T)$ parameter space, or equivalently the $(1+\delta,T)$ space, where $\delta=\rho/\rhob-1$ is the density contrast. The warm IGM is constrained to lie within the white central region in Fig.\ref{diagz0} which is surrounded by exclusion domains. In counter-clockwise order, starting from the right side, these correspond to 1) a fast cooling region where the cooling time $\tcool$ is shorter than the Hubble time $t_{\rm H}$ (so that the gas cannot remain in this region), 2) a high-temperature and high-density domain made of very rare and massive potential wells which are beyond the cut-off of the probability distribution function (pdf) $\cP(\rho)$ (hence they may occur, but with a vanishingly small probability as one moves away from the boundary), 3) linear scales where the gravitational dynamics has not heated the gas yet through shocks, 4) rare voids below the low-density cutoff of the pdf $\cP(\rho)$, and 5) a low-temperature region excluded by radiative heating since the UV background flux heats the gas (here the Lyman-$\alpha$ forest) up to $T \ga 10^3$ K.

The IGM can then be split into two phases. Firstly, a ``cool'' component described by a well-defined equation of state lies along the lower bound set by radiative heating in the range $10^{-1} \la 1+\delta \la 6$. This corresponds to the Lyman-$\alpha$ forest where the temperature is set by the interplay of radiative heating by the UV background with the expansion of the universe. Secondly, a warm component is broadly distributed within the central region, around the curve labeled ``warm IGM'' in Fig.\ref{diagz0}. This gas is shock-heated by the gravitational dynamics, up to the temperature of the large-scale structures which are just turning non-linear. Finally, some of the gas which enters the cooling region forms stars and galactic disks (note that feedback from supernovae and stellar winds only allow a small fraction of this matter to turn into stars, e.g. Valageas \& Schaeffer 1999). The phase-diagram displayed in Fig.\ref{diagz0} agrees with the results of numerical simulations (e.g., Dave et al. 2001). We refer the reader to Valageas et al. (2002) for a detailed discussion of these processes and of their redshift evolution.

\section{Entropy sources}

As recalled in the Introduction, in order to break the $\LX \propto T^2$ rule, which holds (Kaiser 1986) when the cluster gravitational potential is the sole source of energy, we may look for additional heating mechanisms. In this paper, we investigate the gravitational shock-heating induced by the {\it larger scales} which are just turning non-linear. More precisely, we consider the first scenario described in the Introduction, where this transfer of energy from such large scales down to the ICM proceeds via the warm IGM. Thus, the gas is first shock-heated within surrounding filaments and next falls into the cluster potential well. Therefore, we compute below in Sect.\ref{Entropy of the warm IGM} the entropy $\Sw$ of the warm IGM. However, some of the gas within the ICM does not correspond to this late-falling matter. Indeed, part of the ICM observed at $z=0$ was already embedded within virialized structures at $z=1$ (within loose groups or cooled galaxies) and was later on ejected (through ram pressure, supernovae winds, tidal stripping,...) within the diffuse ICM. Hence we compute in Sect.\ref{Entropy associated with substructures} the properties of this second ICM component, ejected from substructures within the cluster. Hereafter, following Ponman et al. (1999), we define\footnote{The thermodynamical specific entropy is thus, within an additional -irrelevant- constant, equal to $3/2 \ln S$.} the specific entropy of the gas as:
\beq
S = \frac{ k T \; \nb^{-2/3}}{1 \; {\rm keV} \; {\rm cm}^2}  ,
\label{S}
\eeq
where $\nb$ is the baryon number density.

\subsection{Entropy associated with the larger scales: the warm IGM contribution}
\label{Entropy of the warm IGM}

From the phase-diagram shown in Fig.\ref{diagz0}, we compute the mean\footnote{To average the entropy over patches of different  entropy $S_i$ but also different densities $n_i$ it is more relevant to bring them adiabatically to the same effective density, say $n_{\rm eff}$ and then take the mass-weighted average $\lag T \rag = \lag {S_i}\rag n_{\rm eff}^{2/3}$ of their rescaled temperature $ {S_i} n_{\rm eff}^{2/3}$ and next to define the average entropy $\lag S \rag$ by eq.(\ref{S}): $\lag S \rag  = \lag T \rag/n_{\rm eff}^{2/3}$. This procedure is independent of the choice of $n_{\rm eff}$ and leads simply as the average specific entropy to the mass weighted average (\ref{Sw}) of $S$ defined in eq.(\ref{S}). A straightforward way to convince oneself that this is the correct average is to see that this procedure provides the correct final entropy in case the patches are physically mixed at the fixed density $n_{\rm eff}$. This choice is the one which is compatible with our final use of $\Tad$ in Sect.\ref{The temperature-luminosity relation of clusters}.} specific entropy $\Sw(z)$ at redshift $z$ of the warm IGM, created by the just-collapsing large scales, as:
\beq
\Sw(z) =     \frac{1}{F_{\rm w}} \int {\d x} \; x h_{\rm w}(x) \; {S(x)}  ,
\label{Sw}
\eeq
with:
\beq
F_{\rm w}(z) = \int {\d x} \; x h_{\rm w}(x) .
\eeq
The entropy $S$ entering the average (\ref{Sw}) is the one defined in eq.(\ref{S}) and the integral runs along the upward rising curve in Fig.\ref{diagz0} labeled ``Warm IGM''. The variable $x \equiv (1+\delta)/\xib$ (where $\xib(R) = \lag \dR^2 \rag$ is the mean two-point correlation over a sphere of radius $R$) is a coordinate which, in eq.(\ref{Sw}), is taken at a given redshift $z$ along the mean equation of state of the warm IGM. The quantity $x h_{\rm w}(x) \d x$ is the mass fraction enclosed within the interval $[x,x+\d x]$ and the mass function $h_{\rm w}(x)$ is proportional to the pdf $\cP(\rho)$ as described in Valageas et al. (2002), normalized to the total mass within the warm IGM component. Thus, $F_{\rm w}$ is the fraction of matter enclosed within the warm IGM phase.

We display our result for $\Sw$ as the solid line in Fig.\ref{Sz}. We see that the entropy $\Sw$ of the warm IGM increases with time, since the mean density of the universe declines while the characteristic temperature grows as larger gravitational structures form. This is due to the increasing depth of the potential well of the typical object which turns non-linear (i.e. the virial temperature of the late forming clusters is higher than that attached to the early forming galaxies). This is also apparent in Fig.1 of Valageas et al. (2002) which gives the time evolution of the phase diagram of Fig.\ref{diagz0}.

\subsection{Entropy associated with the smaller scales: the substructure contribution}
\label{Entropy associated with substructures}

Here we turn to the second component of the ICM: the gas which was already embedded within virialized objects at earlier redshifts. To this order, we first consider the properties of the virialized halos which form at a given redshift $z$.

\subsubsection{Entropy within objects newly formed at redshift $z$} 

We first define as in eq.(\ref{Sw}) the mean entropy $\Svir(z')$ attached to the virialized halos newly formed at redshift $z'$:
\beq
\Svir(z') =    \frac{1}{\Fvir} \int {\d x} \; x h_{\rm vir}(x) \; {S(x)} ,
\label{Svir}
\eeq
with:
\beq
\Fvir(z')=  \int {\d x} \; x h_{\rm vir}(x) .
\label{Fvir}
\eeq
The quantity $x h_{\rm vir}(x) \d x$ is again the mass fraction enclosed within the interval $[x,x+\d x]$, but the integration now runs along the line $\delta=\Deltavir$, above the lower branch of the cooling curve. This is shown by the vertical dotted line in Fig.\ref{diagz0}, see also Fig.1 in Valageas et al. (2002). The density contrast threshold $\Deltavir \sim 200$ is obtained from the usual spherical model. In particular, $\Fvir(z')$ is the fraction of matter embedded within these virialized halos at a redshift $z'$. In eq.(\ref{Fvir}) the lower bound of the integration is set by the lower branch of the cooling curve. This quantity was already computed in Valageas et al. (2002) (solid line in Fig.3 there) and checked against numerical simulations. To obtain the entropy $S(x)$ of a given halo we use the definition (\ref{S}) at the adequate density $n_b$, setting $T=\Tvir$, where $\Tvir$ is the virial temperature:
\beq
\Tvir = \frac{\cG \mu m_p M}{2 k R} ,
\label{Tvir}
\eeq
$M$ is the mass and $R$ the radius of the halo. Therefore, the mean $\Svir$ describes the mean entropy of just virialized halos formed at a given redshift $z'$, generated during their own  gravitational collapse (whence the use of $\Tvir$).

Although we integrate eq.(\ref{Fvir}) along the density threshold $\Deltavir$ at redshift $z'$, it allows us to describe the high-density substructures present within virialized clusters at later redshifts $z<z'$. Indeed, as noticed in Navarro et al. (1996) the core density within dark matter halos roughly scales as the density of the universe at the redshift when this scale turned non-linear. This scaling is also consistent with the stable-clustering {\it Ansatz} (Peebles 1980). Therefore, the matter described by eqs.(\ref{Svir})-(\ref{Fvir}) at redshift $z'$ actually corresponds to substructures with a density contrast $\sim \Deltavir(z') (1+z')^3$ at redshift $z=0$ within larger just-virialized cluster, see Sect.~\ref{Entropy of a collection of condensed objects} below.

We estimate the entropy available through ejection of mass from these virialized objects by this specific entropy $\Svir(z')$. Indeed, the gas which can escape from smaller-scale halos (e.g., galaxies) must have a kinetic or thermal energy above the potential well of these objects (e.g., the velocity of outflows must be larger than the local escape velocity). We assume that the additional energy poured into the ICM scales as the potential energy of this subunit. Here, the advantage of using the entropy as a basic variable is that shock-heating which may occur after virialization (when small-scale objects feed the intermediate scale structures) can only increase the entropy of the system. Moreover, it takes into account the expansion of the gas as it spreads within the cluster.

\subsubsection{Entropy of a collection of condensed objects formed prior a given epoch $z$}
\label{Entropy of a collection of condensed objects}

The substructures located within a new larger-scale cluster were formed in the past so that there is still an average to be made over their respective epoch of formation. This mean is expected to be dominated by contributions $\Svir(z')$, at definitely earlier times $z' > z$, and $\Svir(z)$ enters this sum only marginally.

\begin{figure}
\psfig{figure=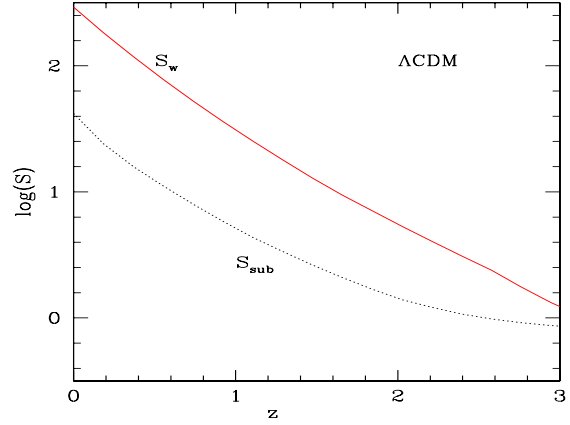,width=8cm,height=6cm}
\caption{The redshift evolution of the characteristic entropies. The solid curve $\Sw(z)$ corresponds to the mean specific entropy acquired by warm IGM gas from scales above the cluster scale at redshift $z$. The dotted line $\Ssub$ corresponds to the mean specific entropy of the baryons within all substructures in a given cluster which virialized before redshift $z$.}
\label{Sz} 
\end{figure}

We now proceed evaluating this entropy, $\Ssub$, within substructures existing at redshift $z$, averaged over their previous redshift of formation and weighted by the mass flux entering into virialized objects. More precisely, we write for the average entropy $\Ssub$ of these structures:
\beq
\Ssub(z) = \frac{1}{\Fvir(z)} \int_z^{\infty} \d z' \; \frac{\d \Fvir}{\d z}(z') \; {\Svir(z')}  .
\label{Ssub}
\eeq
The meaning of $\Ssub$ is the following. A new cluster builds up at redshift $z$ as a large scale density fluctuation turns non-linear and older smaller-scale structures gather within this deeper potential well. Then, these substructures, already virialized within the just-forming cluster of interest, may also be the source of entropy for this new cluster. Indeed, the gas embedded within such smaller objects (e.g., galaxies) may escape from these subunits and become part of the intra-cluster medium (ICM), for any reason (supernovae feedback, tidal stripping, mergers), so it can bring some entropy to the more diffuse ICM. 

We display $\Ssub(z)$ as the dotted line in Fig.\ref{Sz}. As expected, $\Ssub(z)$ declines at higher redshift, as  $\Sw$, because at earlier times typical potential wells were shallower. The average over redshift which enters the definition (\ref{Ssub}) also implies that $\Ssub(z)$ is significantly smaller than $\Svir$. Thus, we obtain numerically $\Ssub(z=0) \sim \Svir(z=1)$ which means that an important fraction of the matter which enters virialized halos at $z=0$ was already embedded within collapsed structures at $z \sim 1$ (note that this is consistent with numerical simulations, e.g. Borgani et al. 2002). So, the major contribution to $\Ssub$ at the present epoch may be attributed to objects which form at $z \sim 1$, that somewhat loosely one may call galaxies. But it should be clear that in our average specific entropy {\it all} substructures are accounted for.

\subsection{Relative efficiency of external energy sources}
\label{Relative efficiency of external energy sources}

The entropy $\Ssub$ may provide an additional contribution to the cluster entropy floor, independently of $\Sw$. However, this entropy is initially ``trapped'' within the substructures, and we now discuss to what extent it is available to the intracluster gas. Some feedback from the gas trapped within galaxies, for instance, is expected, since the cluster gas is known to contain metals. But it is also expected that galaxies retain a non-negligible fraction of their gas.  As an example to start with, we shall adopt in Sect.\ref{The temperature-luminosity relation of clusters} the simple rule that this is the source of about $1/3$ of the cluster gas, an estimate which is consistent with the typical metallicity, $1/3$ solar, of a cluster.
Thus, we may estimate the external entropy level to be
\beq
\Sex = \left( \frac{1}{3} {\Ssub} + \frac{2}{3} {\Sw} \right) .
\label{Sex}
\eeq
This is the essence of hierarchical clustering, with a contribution from the larger just-collapsing structures as well as from the smaller, already collapsed objects, the dark matter as well as the attached baryons  being incorporated into the newly forming structures.

Note that both processes, leading to either $\Sw$ or $\Ssub$, break the simple scaling laws which led to $\LX \propto T^2$ and, if sufficiently strong, might explain the observed temperature-luminosity relation of clusters (as with other physical processes based on supernovae or quasar-driven energy injection which also yield an entropy floor). First, the formation of large-scale structures through gravitational instability builds a network of filaments and shock-heated regions which provide a background entropy and temperature for the gas which eventually gathers to build a cluster. Second, a collapsing group has inner substructures which may be relevant to its subsequent evolution. 

These entropy floors however have a negligible impact on the formation of very rare and massive clusters, characterized by high temperatures. This may also be tied to the fact  that rare events are known to be almost spherically symmetric (e.g., Bernardeau 1994, Valageas 2002), while the ones under the influence of external heating by the larger scale filaments are precisely those which are expected not to bear the spherical symmetry. Most important, let us note that our new heating process which we have assumed to contribute to structures ranging from rich clusters (where it turns out to be subdominant) down to small groups (where it modifies significantly the gas distribution of the whole cluster) needs only to be present in the latter for our explanation for the breaking of the $\LX \propto T^2$ rule to hold.

\section{The temperature-luminosity relation of clusters}
\label{The temperature-luminosity relation of clusters}

We compute the temperature - bolometric luminosity relation of clusters as in Valageas \& Schaeffer (2000) (see Sect.5). Hence, we assume approximate hydrostatic equilibrium for the gas within a dark matter halo described by the density profile:
\beq
\rho(r) = \frac{\rho_c}{(r/r_s)^{1.5} \left[ 1 + (r/r_s)^{1.5} \right]} 
\hspace{0.4cm} \mbox{with} \hspace{0.4cm} r_s=\frac{R}{c} ,
\label{Moore1}
\eeq
where $c$ is the concentration parameter and $R$ the virial radius. The latter is defined as usual as the radius where the overall density contrast reaches the threshold $\Deltavir \sim 200$ given by the standard spherical model. We use $c=4$ as in Moore et al. (1999), although this parameter may depend slightly on the mass of the cluster. Besides, at each radius $r$ we can associate an effective temperature $\Ts(r)$ which measures the depth of the potential well, defined by:
\beq
\Ts(r) = \frac{{\cal G} \mu m_p M(<r)}{2 k r} = T \; \frac{ \ln \left( 
1+(r/r_s)^{1.5} \right) }{ \ln \left( 1+c^{1.5} \right) } \; \frac{R}{r} .
\label{Tsr}
\eeq
Thus, $\Ts(r)$ yields the temperature which can be reached by the gas through non-adiabatic shock-heating during the formation of the cluster. On the other hand, since the initial entropy (or temperature) of the gas is non-zero we can also define the temperature $\Tad(r)$ associated with an adiabatic compression during the infall, from the typical density of the warm IGM (i.e. $\delta \sim 10$) up to the density $\ng(r)$ reached within the cluster (i.e. $\delta > \Deltavir$). It is set by the entropy floor $\Sex=\Sw$ through the relation (see eq.(\ref{S})):
\beq
\Tad(r)= \frac{\ng(r)^{2/3}}{1 \; {\rm cm}^{-2}} \; {\Sex} \; {\rm keV} ,
\label{Tad}
\eeq
since it corresponds to a constant-entropy (i.e. adiabatic) dynamics. Thus, $\Tad$ is the temperature reached by some adiabatic gas falling from the outer warm IGM into the cluster, with an initial entropy $\Sex$ acquired in a larger-scale filament. In order to take into account the fact that distinct regions of the warm IGM can have different entropies, within the range allowed by the phase-diagram shown in Fig.\ref{diagz0}, we consider that the gas thermalizes within the cluster at a typical density $\delta \sim \Deltavir$. From eq.(\ref{Tad}), we see that this amounts to using as a unique initial entropy level the mean $\Sw$ defined in eq.(\ref{Sw}). Moreover, through eq.(\ref{Sex}) this also allows us to take into account the gas which did not come lately from the warm IGM but was ejected from substructures. Next, in order to take into account both these possible external sources and the shock-heating associated with the cluster itself, assuming all sources of entropy correspond to matter which is well-mixed throughout the whole cluster
\footnote{We indeed expect the small, somewhat irregular, groups where this contribution is most important to have undergone substantial mixing.}, we write:
\beq
\Tg(r) = \Ts(r) + \Tad(r) .
\label{Tgas1}
\eeq
Thus, if $\Ts > \Tad$, the physics of the gas is governed by the local potential well and $\Tg \simeq \Ts$. This corresponds to massive high-temperature clusters or to the outer shells. On the other hand, if $\Ts < \Tad$ the behaviour of the gas is governed by the external entropy source. In fact, the density of the gas $\rhog$ then shows a plateau within a core radius $\Rcore$ where both terms in eq.(\ref{Tgas1}) are equal. Indeed, when $\Tg \gg \Ts$ the gas no longer falls into the potential well and a core with a nearly constant baryon density appears. On the other hand, at larger radii the gas follows the dark matter and we take $\rhog=\Om/\Ob \rho$. Thus, for smaller halos the preheating becomes more important and the gas density profile gets flatter, decreasing the X-ray luminosity. We refer the reader to Valageas \& Schaeffer (2000) for a more detailed discussion of this model (and the role of radiative cooling).

\begin{figure}
\psfig{figure=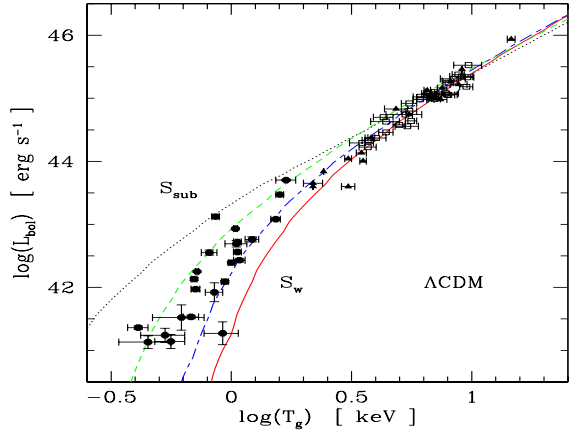,width=8cm,height=6cm}
\caption{The temperature - bolometric luminosity relation of clusters at $z=0$, using $\Sex=\Sw$ (solid line), $\Sex=\Ssub$ (dotted line) or the mean $\Sex$ defined as in eq.(\ref{Sex}) (lower dashed line). The upper dashed line corresponds to the lowest acceptable entropy injection which can account for the data. The latter are taken from Helsdon \& Ponman (2000) (filled circles), Arnaud \& Evrard (1999) (triangles) and Markevitch (1998) (squares).}
\label{LXT} 
\end{figure}

Finally, we display in Fig.\ref{LXT} the $\LX-T$ relation of clusters obtained in this way at $z=0$. The solid line is $\Sex=\Sw$, the dotted line $\Sex=\Ssub$ while the lower dashed line is the mean defined in eq.(\ref{Sex}). We recover the knee at low luminosities that is found in observations. This feature is common to all models which include some preheating of the gas which provides an entropy floor of order $100$ keV cm$^2$. However, we see that the case $\Sex=\Ssub$ provides an entropy level which is too small. Therefore, gas escaping from older substructures cannot explain the extra entropy of small groups. In particular, if the only source of entropy were galactic outflows (driven by supernovae or any other process), this would not suffice to account for the observed entropy floor. This is consistent with detailed studies of preheating by supernovae and stellar winds (e.g., Valageas \& Silk 1999b, Wu et al. 2000). The warm IGM, on the other hand, provides an important contribution. In particular, we see that even in the case defined by eq.(\ref{Sex}), where the energy brought by the warm IGM is diluted by a factor $2/3$ (i.e. only $2/3$ of the current X-ray emitting gas came from the outer IGM over a redshift range $\Delta z \ll 1$), the entropy floor is sufficient to explain the knee of the $\LX-T$ relation. An interesting exercise is to consider the lowest possible entropy injection required by the data (left dashed-line in Fig.\ref{LXT}). This may be achieved even with no entropy contribution from galactic outflows provided $38\%$ of the cluster gas originates from the recently formed warm IGM (the other $62\%$ having a negligible entropy). If we include our (modest, but probably quite realistic) estimate of the entropy associated with galactic outflows the requirement is that at least $28\%$ of the ICM bears the entropy of the warm IGM phase (the other $72\%$ bearing $\Ssub$). These modest requirements are quite plausible, indeed, and strongly back our suggestion.

This entropy floor agrees with the plateau associated with small groups (e.g., Lloyd-Davies et al. 2000). Thus, in this case the properties of the warm IGM would explain in a natural fashion the observed properties of groups and clusters. From the $\LX-T$ relation we can also compute the luminosity function of clusters and its evolution with redshift. Such a study is described in Valageas \& Schaeffer (2000).

We should emphasize that the underlying picture of our model is that the entropy floor is connected with the warm IGM, and to its equation of state which involves regions of density contrast $\delta \sim 10$ and scales much larger than the cluster scale. This is quite different from the alternative suggestion of Voit \& Bryan (2001). In the latter model, for instance, it is seen from their Fig.1 that their entropy threshold is associated with the internal cooling of the cluster, in very high density regions of $\delta \sim 10^4$, namely those actually observed in the centres of clusters which have {\it completed} their formation process. Their suggestion actually relies on a very specific feature of the cooling curve, which, drawn in the $S(T)$ variables, happens to exhibit a constant-entropy plateau in a very specific -and extremely narrow- range of temperatures. Of course, in our scenario the cooling curve $\tcool=t_{\rm H}$ also comes into play, but as a whole as it defines the high-density boundary of the ``warm'' IGM phase in Fig.\ref{diagz0}. However, the entropy of the gas is not at all defined by the edge of this exclusion region (and hence does not depend on its local behaviour in a small region of phase-space) since the warm IGM component is actually spread over a broad band within the white central region drawn in Fig.\ref{diagz0}, as shown by numerical simulations (e.g., Dave et al. 2001). Besides, the gas is truly preheated before the cluster forms while it is embedded within the complex network of filaments at $\delta \sim 10$. 

Here we may also note that Muanwong et al. (2001) find that their numerical simulations which involve radiative cooling indeed recover a strong increase of the entropy of the X-ray gas within clusters which allows them to match the observed $\LX-T$ relation. However, they remark that the mass of cold gas they obtain is too large. Here we did not explicitly model this process but it simply means that some of the gas which has fallen into the cooling region has been reheated by some process (supernovae being one of the possibilities). This gas may then either remain associated with galactic halos (in which case it is irrelevant for our purpose here) or be ejected into the ICM. This corresponds to the contribution $\Ssub$ discussed in Sect.\ref{Entropy associated with substructures}.

\section{Shock-heating of the ICM itself from larger scales}
\label{Shock-heating of the ICM itself}

Finally, we briefly discuss here the second scenario we presented in the Introduction. This corresponds to the case where the gravitational shock-heating associated with the large scales which are just turning non-linear directly heats the ICM {\it in situ}, without involving the transfer of matter from the warm IGM. The characteristic temperature involved by this scenario is again the temperature (or shock velocity) associated with the non-linear length scale, which is of order $T \sim 1$ keV at $z=0$ (e.g., Cen \& Ostriker 1999, Valageas et al. 2002). Therefore, such shock-heating may indeed influence the properties of groups with $T < 1$ keV. In fact, this is the same ``coincidence'' which underlies the scenario we described in the previous sections, which relies on the warm IGM as an intermediate tool to realize this energy transfer from the non-linear length scale down to the ICM scale. However, in order to increase the entropy of the ICM by the same level as the first scenario this model requires a stronger shock-heating because of the higher density of the gas within the cluster as compared within filaments, see eq.(\ref{S}). Indeed, we can see from Fig.1 that by heating the gas within filaments we do not need to systematically reach high temperatures $T \sim 1$ keV in order to achieve high entropies, because of the smaller density of the warm IGM (the density contrast $\delta$ typically runs from 6 to 200). On the other hand, although large-scale shocks (which may extend over $1-5 h^{-1}$ Mpc) are indeed seen in numerical simulations to envelop and penetrate deep inside clusters (e.g., Miniati et al. 2000) it is difficult to estimate their efficiency. We shall not investigate this scenario further in this work but we expect it can provide an alternative way of injecting energy into clusters from larger scales.

\section{Conclusion}
\label{Conclusion}

We have shown that the entropy of the warm IGM can be sufficient to explain the observed entropy of small clusters. This scenario relies on the assumption that the gas which has recently falled into the cluster from the surrounding filaments has been efficiently mixed within the halo so as to form at least $28 \%$ of the ICM at the radii where the extra-entropy is observed. This point is clearly a matter of debate since it is not obvious to follow the behaviour of the gas within such complex objects as clusters. However, we feel this requirement is quite plausible. A natural scale appears in our scenario, $1$ keV, the potential energy of the scales just turning non-linear. Clusters below $1$ keV where this external entropy source can play a role are specifically the ones under debate. These rather small clusters should be rather irregular so that their accretion is not expected to be well-ordered or spherically symmetric. Then, the late gas coming from the warm IGM may indeed spread over the whole group, taking advantage of the general mixing (e.g., turbulence) brought by the chaotic dynamics of the various components (e.g., galaxies) of the group. On the other hand, we briefly notice that one could also imagine a second scenario. There, the gravitational energy associated with the large scales which are just turning non-linear could directly heat the ICM itself. This assumes that shocks generated by those large scales can spread within the cluster and heat at least $28 \%$ of the ICM. However, in order to reach the same entropies this scenario requires a stronger shock-heating than in the first model because of the need to compensate for the higher density of the gas within the ICM as compared with filaments.

Altough the efficiency of these mechanisms remains to be evaluated with a better accuracy before one can safely determine whether they indeed play an important role in the physics of clusters, we note that such scenarios present the advantage to dispense with the need of a new external preheating source (e.g., supernovae or quasars). Indeed, they only rely on gravitational shock-heating which has recently been seen to explain the properties of the IGM in numerical simulations (e.g., Dave et al. 2001). Therefore, we merely propose here that the mechanisms which build the warm IGM may also govern the properties of small clusters. So far, numerical simulations have not really substantiated this claim since they usually recover the standard scaling $L_X \propto T^2$, unless one adds some ad-hoc preheating or takes into account cooling processes (which lead to other problems like overcooling). However, they may not have been able to pin down all salient effects of shock-heating mechanisms yet. 

Similar processes may be expected to be at work at the epoch of galaxy formation. They may for instance modify the hydrodynamical equilibrium of the smaller galaxies. In particular, as already noticed in Valageas \& Silk (1999b) preheating models are strongly constrained by the contradictory requirements arising from the observed properties of clusters and galaxies. This will be studied in a forthcoming paper for the process presented here which exhibits a specific evolution with redshift.

In summary, there are three hypothesis for the entropy floor in a given cluster. It may arise 
1) from external preheating by supernovae or quasars;
2) from a very specific feature of the cooling curve, as implied by the suggestion of Voit \& Bryan (2001);
3) as we suggest here, from the pre-collapse phase of structures much larger than clusters which transfer their energy to the smaller cluster scales through shock-heating, which also builds the warm IGM phase. 

We have stressed the difficulties encountered by the two first possibilities.
The advantage of the latter explanation is that it is a natural outcome of hierarchical scenarios of large-scale structure formation. 
Numerical simulations and observations may be used to check the entropy evolution implied by our suggestion.

\section*{acknowledgements}

We are very grateful to the referee, Trevor Ponman, for his stimulating comments, helping us to improve this paper.

\end{document}